%% file: main.tex
\documentclass[10pt,letterpaper,conference]{IEEEtran}
\IEEEoverridecommandlockouts

\usepackage{cite}
\usepackage{amsmath}
\usepackage{amssymb}
\usepackage{adjustbox}
\usepackage{algorithmic}
\usepackage{fixltx2e}
\usepackage{graphicx}
\graphicspath{ {images/} }

\usepackage{kbordermatrix}

\usepackage{enumerate}

\usepackage{url}
\usepackage{algorithmic}
\usepackage{subfigure}
\usepackage{algorithm}
\usepackage{multirow}
\newlength\myindent
\usepackage{tikz}
 \usetikzlibrary{shapes, arrows, decorations, decorations.markings, backgrounds, calc, arrows.meta, shadows, shadows.blur,positioning,spy,shadings}
\pgfdeclarelayer{myback}
\pgfdeclarelayer{myback2}
\pgfsetlayers{background,myback2,myback,main} 

\usepackage{pgfplots}
\pgfplotsset{width=10cm,compat=newest}
\pgfplotsset{compat=newest} 
\pgfplotsset{plot coordinates/math parser=false} 
\newlength\figureheight 
\newlength\figurewidth 
\usepgfplotslibrary{units}
\usepackage{pgfplotstable}

\tikzstyle{vecArrow} = [{line width=0.5pt, double distance=5pt, arrows={-Implies[length=0pt 0.8 0]}}]
\tikzstyle{vecNoArrow} = [anchor=center,decoration={markings,mark=at position
	1 with },
double distance=5pt, shorten >= 3pt,
preaction = {decorate},
postaction = {draw,line width=1.5pt, white,shorten >= 1.5pt}]
\tikzstyle{vecArrowTUMblue} = [anchor=center,decoration={markings,mark=at position
	1 with {\arrow{Straight Barb[length=6pt,TUMblue]}}},
double distance=5pt, shorten >= 3pt,
preaction = {decorate},
postaction = {draw,line width=1.5pt, white,shorten >= 1.5pt}]
\tikzstyle{innerWhite} = [semithick, white,line width=4pt, shorten >= 4.5pt]
\tikzset{
	triangle/.style = { regular polygon, regular polygon sides=3, rotate=-90,thick,scale=1.2},
	triangleWithText/.style = { regular polygon, regular polygon sides=3, shape border rotate = -90, rotate=0,thick,scale=1.2},
	sum/.style = {draw, circle,thick,scale=1.3},
}

\definecolor{mycolor1}{rgb}{0.00000,0.44700,0.74100}%
\definecolor{mycolor2}{rgb}{0.85000,0.32500,0.09800}%
\definecolor{mycolor3}{rgb}{0.92900,0.69400,0.12500}%
\definecolor{mycolor4}{rgb}{0.49400,0.18400,0.55600}

\usepgflibrary{patterns} 
\usepgflibrary[patterns] 
\usetikzlibrary{patterns} 
\usetikzlibrary[patterns] 

\usepackage{textcomp}

\newcommand\copyrighttext{%
	\footnotesize \textcopyright 2018 IEEE.  Personal use of this material is permitted.  Permission from IEEE must be obtained for all other uses, in any current or future media, including reprinting/republishing this material for advertising or promotional purposes, creating new collective works, for resale or redistribution to servers or lists, or reuse of any copyrighted component of this work in other works.
}
\newcommand\copyrightnotice{%
	\begin{tikzpicture}[remember picture,overlay]
	\node[anchor=south,yshift=10pt] at (current page.south) {\fbox{\parbox{\dimexpr\textwidth-\fboxsep-\fboxrule\relax}{\copyrighttext}}};
	\end{tikzpicture}%
}

\newcommand{\journal}[1]{#1} 
\renewcommand{\journal}[1]{} 

\large{\title{Reinforcement Learning Scheduler for Vehicle-to-Vehicle Communications Outside Coverage}}

\author{
	\IEEEauthorblockN{
			Taylan \c{S}ahin\IEEEauthorrefmark{1}\IEEEauthorrefmark{2}, 
			Ramin Khalili\IEEEauthorrefmark{1},
			Mate Boban\IEEEauthorrefmark{1} and
			Adam Wolisz\IEEEauthorrefmark{2}
		}

		\IEEEauthorblockA{
			\IEEEauthorrefmark{1}Huawei Technologies German Research Center, 80992 Munich, Germany\\
			Email: \{taylan.sahin, ramin.khalili, mate.boban\}@huawei.com
		}
		
		\IEEEauthorblockA{
			\IEEEauthorrefmark{2}Telecommunication Networks Group, Technische Universit{\"a}t Berlin, 10587 Berlin, Germany\\
			Email: adam.wolisz@tu-berlin.de
		}
}

\begin{document}
	\maketitle
	\copyrightnotice
	
	\input{Abstract}
	
	\input{Introduction}
	
	\input{System_Model}

	\input{Proposed_Algorithm}

	\input{Results}

	\input{Conclusions}

	\input{Acknowledgment}

	\bibliographystyle{IEEEtran}
	\input{main.bbl}

\end{document}

%% file: Abstract.tex
\begin{abstract}

Radio resources in vehicle-to-vehicle (V2V) communication can be scheduled either by a centralized scheduler residing in the network (e.g., a base station in case of cellular systems) or a distributed scheduler, where the resources are autonomously selected by the vehicles. The former approach yields a considerably higher resource utilization in case the network coverage is uninterrupted. 
However, in case of intermittent or out-of-coverage, 
due to not having input from centralized scheduler, vehicles 
need to revert to 
distributed scheduling. 

Motivated by recent advances in reinforcement learning (RL), we investigate whether a centralized learning scheduler can be taught to efficiently pre-assign the resources to vehicles for out-of-coverage V2V communication. 
Specifically, we use 
the actor-critic RL algorithm to train the centralized scheduler to provide non-interfering resources to vehicles \emph{before} they enter the out-of-coverage area.




Our initial results show that a RL-based scheduler can achieve performance as good as or better than the state-of-art distributed scheduler, often 
outperforming it. Furthermore, 
the learning process completes within a reasonable time 
(ranging from a few hundred to a few thousand epochs), thus making the RL-based scheduler a promising solution for V2V communications with intermittent network coverage.

\end{abstract}


%% file: Introduction.tex
\section{Introduction}
\label{Introduction}

Vehicle-to-everything (V2X) communication aims at enabling safer and more convenient driving, at the same time improving road capacity. 
For many use-cases defined in~\cite{3gppTS22186}, the most efficient way for vehicles to exchange information is via direct vehicle-to-vehicle (V2V) communications: e.g., sending safety-critical messages containing their position and velocity, or exchanging messages within a platoon. 

Reliability of V2V communications depends in large part on the resource allocation and scheduling. 
Resource allocation in V2V involves a large number of vehicles that request resources, the number of which is also usually large. Combined 
with a range of constraints imposed by the requirements of the applications that the vehicles run, it becomes a difficult online decision making task. Furthermore, compared to other types of networks, V2X network has a highly mobile  environment. This makes the problem of resource allocation even more challenging. 

The most reliable approach to scheduling V2V messages is through a 
centralized scheduler, which has control over the access of vehicles to the radio resources, in order to ensure a reliable V2V communication~\cite{3gppTR36885,3GPPTS36300}. However, in case of intermittent network deployment, there will exist situations where the vehicles do not have connection to a centralized scheduler. 
Similarly, 
vehicles may travel through areas where connection to the network infrastructure is physically impeded (e.g., tunnels). 
In such cases, up to now the only viable option was to revert from a centralized to decentralized scheduling~\cite{3gppTR36885,Gozalvez}. Since the decentralized scheduling algorithms use only a limited input made available by the centralized scheduler (e.g., resource pool assignments), with the remaining information (e.g., traffic demands of vehicles not in vicinity, interference from far-away transmitters that could create hidden node problem, etc.) unused, and given that centralized scheduler cannot provide immediate assignments in the out-of-coverage (OOC) areas, we explore how a centralized scheduler could \emph{pre-schedule} the resouces to vehicles for OOC.

In our previous work~\cite{sahin2018radio}, we explored potential performance that a centralized V2V scheduler could have in cases of OOC, wherein the scheduler pre-schedules the resources for periodic services and reserves the resources for non-periodic services. We explored how the effect of errors in terms of vehicle speed and radio propagation impact the performance of the scheduler. 
In this paper, motivated by recent work showing the potential of reinforcement learning (RL) on network resource management problems in general~\cite{mao2016resource}, and V2V resource management in particular~\cite{magazine}, we implement 
a centralized scheduling algorithm that learns by using the information available in V2X environment, such as the occupancy of radio resources. More specifically, we use the actor-critic deep RL algorithm~\cite{sutton1998reinforcement} to assign the available resources for \emph{OOC periodic transmissions} to the vehicles \emph{before they exit the network coverage}. We evaluate the performance of the algorithm against state-of-art 3GPP solutions for OOC scheduling~\cite{3gppTR36885}, and show that in environments and scenarios varying from simple to complex, and from under-loaded to over-loaded, including half-duplex and realistic channel conditions, 
the proposed algorithm outperforms current solutions.

\subsection{Related Work}
Cellular communication standard 3GPP LTE-A Release 14 is among the current wireless technologies providing support for V2X services, also known as LTE vehicular (LTE-V)~\cite{3gppTR36885}. Owing to the direct radio interface between the vehicles, called as sidelink (SL), V2V communication is supported both in coverage and OOC situations. In terms of radio resource management, two modes exist for SL V2X. In coverage, SL transmission resources can be scheduled by the cellular infrastructure, i.e., base stations (BSs), in a centralized way, referred to as ``Mode-3''. On the other hand, especially in OOC, vehicles are allowed to autonomously select the resources they transmit using a sensing-based mechanism referred to as ``Mode-4''. Resource pools can be (pre-)configured for both modes, constraining the set of time and frequency resources each mode can use, also optionally based on geo-location of the vehicles, mainly to manage SL interference.



Performance of SL V2X has been studied by the industry and the academia under various use cases and assumptions (refer to e.g., \cite{Gozalvez}, \cite{GozalvezCentralized}, and \cite{cv2x11p}). However, so far, application of machine learning (ML) to any resource allocation problem targeting V2X is in a nascent phase. An overview on the state of the art of applying ML to the challenges of vehicular networks, including the resource management is provided in \cite{magazine}. The authors of~\cite{magazine} point out that the highly dynamic nature vehicular networks challenges the conventional methods for resource management. 
Instead, RL could be an alternative effective solution, which interacts with, and adapts its actions to the unknown environment. A distributed resource allocation mechanism based on deep RL is proposed by the same authors in \cite{RLv2v} and \cite{RLv2v2}, where the challenges of satisfying the latency constraints of broadcast V2V messages, and minimizing the interference to vehicle-to-infrastructure (V2I) links are tackled. The vehicular wireless standard IEEE 802.11p is modified with RL in~\cite{RL11p}, with the aim of overcoming the scaling problem with the increased vehicular density. 

In terms of centralized scheduling algorithm for V2X networks, \cite{RLv2i} applies RL to scheduling of non-safety downloads over V2I links, whereas \cite{sdn} and \cite{cloud} deal with the management of virtual resources (including communication, computation, and storage resources) on a software-defined vehicular network using stochastic learning, and applying RL on vehicular cloud, respectively.

To the best of our knowledge, ML for a centralized scheduler managing the resources of V2V communications has not been treated in the literature yet. Motivated by the aforementioned works showing the performance benefits of RL on resource allocation problems, we are interested to observe if and how it could be useful to overcome the stringent requirements of V2X use cases outside coverage, such as reliability and latency. In this study, we perform an exploratory study using several relevant V2V scenarios to investigate if a centralized scheduler can learn to perform resource (pre-)scheduling reliably. 



\begin{figure}[!t]
	\centering
	\includegraphics[width=\columnwidth, height=4cm]{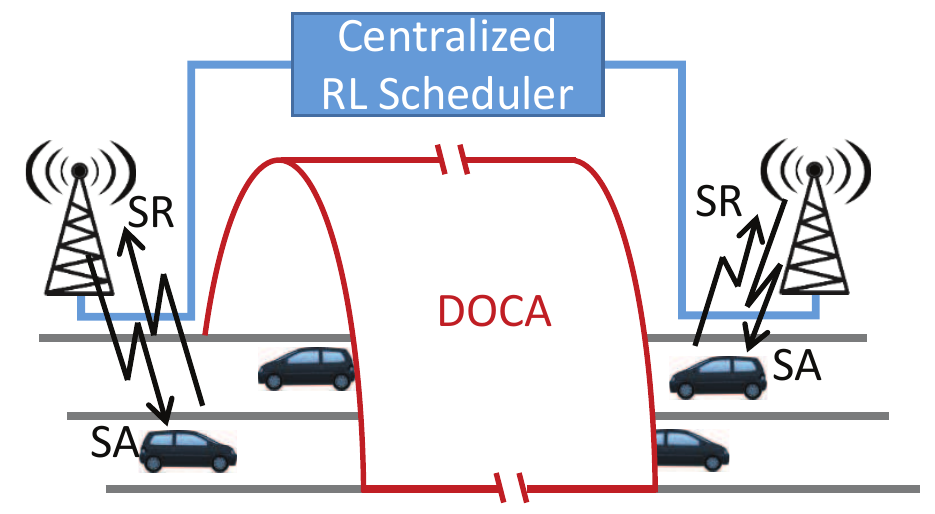}
	\caption{DOCA delimited by BSs on a two-way highway segment, in which vehicles communicate on the resources requested by sending SR and receiving SA before they enter it.}
	\label{doca}
\end{figure}

\subsection{Our Contribution}


We consider an area outside the cellular coverage where vehicles would like to perform V2V communications. This area is  surrounded by the network infrastructure, e.g., delimited by BSs as illustrated in Fig.~\ref{doca}, which we call DOCA (Delimited Out-of-Coverage Area). Our approach differs from the state of the art in two aspects: 
\begin{itemize}
\item We perform the resource (pre-)scheduling for OOC V2V communication using the network infrastructure, based on the information readily available to the infrastructure \emph{while a vehicle is still connected to the network}. Specifically, we exploit the information about the radio resource occupancy, along with the scheduling requests (SRs) for periodic transmission that last a certain period of time before and during the time a vehicle is traversing the DOCA. Vehicles are then informed about the scheduling decision via the scheduling assignments (SAs) before they enter DOCA.
\item To exploit V2X-specific information for resource pre-allocation, we resort to RL, which was shown to apply well to a wide range of problems, such as games involving large combinatorial space, image recognition, and robot movement~\cite{sutton1998reinforcement}, and was recently applied to resource scheduling in vehicular networks~\cite{magazine}. Specifically, we employ a state-of-the-art deep RL, asynchronous-advantage actor-critic algorithm, referred to as A3C algorithm, which was proven to have strong convergence properties~\cite{Mnih2016}.
\end{itemize}

Our results show that, in case of periodic broadcast V2V transmissions in a limited size DOCA, the proposed centralized deep RL scheduler converges to a near-optimal solution, and 
outperforms the state-of-the-art distributed scheduling algorithms with respect to reliability.


The rest of this paper is organized as follows. In Section \ref{Model} we provide our system model and define the problem. RL and its application to our problem are described in Section \ref{Algorithm}. Section \ref{Results} presents the results of our evaluations. Finally, Section \ref{Conclusion} concludes the paper, and discusses the further related work.

%% file: System_Model.tex
\section{System Model}
\label{Model}

We consider vehicles traveling on a highway 
in two directions, as depicted in Fig.~\ref{doca}. V2V application running at each vehicle generates periodic messages to be broadcast to all other vehicles in its vicinity. Messages can be considered as cooperative awareness messages (CAMs) that carry vehicle-specific information such as position and velocity. Vehicles perform V2V communication using the radio resources allocated by the network. Each message, depending on its size, requires a certain amount of time and frequency resources to be transmitted, called a transmission block (TB). Based on the LTE assumptions~\cite{3gppTR36885}, a TB occupies a single time slot called as subframe, and one frequency slot called a subchannel, on the assumed radio resource grid, as illustrated in Fig.~\ref{Pool}.

\subsection{Problem Definition}

We consider the V2V application at each vehicle has a certain communication reliability requirement. This requirement is considered to be fulfilled when transmitted messages are successfully received by a certain target ratio of the vehicles that are within a certain range around the transmitter. We target to ensure the reliability requirement of the V2V application by means of allocating sufficient radio resources for the transmissions of the vehicles. Our performance goal is to ensure the reliability of the periodic broadcast V2V messages in DOCA, measured by the packet reception ratio (PRR, the ratio of successful receivers over the total number of receivers in a certain range). 

For performance evaluations, we consider two different vehicular network environments:

\begin{enumerate}[E1)]
	\item A DOCA of a single collision domain, without any pathloss, that is, all vehicles are within the transmission range of each other.
	Given these conditions, reception of a message is successful if no other transmission takes place on the same radio resource scheduled (i.e., no collision), and the receiver is not scheduled to transmit at the same time, as imposed by the half duplex (HD) constraint.
	\item A DOCA of multiple collision domains, where pathloss and fading effects are taken into account. Hence, successful reception of a message requires the signal-to-interference-plus-noise ratio (SINR) at the receiver to be larger than a certain target level, which depends on the distance between the transmitter and the receiver, as well as the interference level from other transmissions using the same radio resource, besides the half-duplex constraint. Reusing the same radio resource is possible, when the transmitters are sufficiently far from each other so that the SINR does not drop below the target level.
\end{enumerate}

E1 helps us to identify the scheduler performance, and the ability of RL to avoid HD constraint and assigning interfering resources, while abstracting the effects of channel conditions. Whereas, E2 enables a realistic evaluation of our proposed RL scheduler.

%% file: Proposed_Algorithm.tex
\section{Deep Reinforcement Learning Scheduler} 
\label{Algorithm}

We design a learning scheduler 
that manages the V2V radio resources for DOCA, so as to ensure the target reliability required by the V2V applications. The scheduler assigns resources to each vehicle \emph{before} it enters DOCA; the resources will be used by that vehicle throughout its travel in DOCA. 
 Therefore, the scheduler is assumed to be aware of the vehicles entering and exiting DOCA via the BSs located at each end of the DOCA (e.g., along the highway), as illustrated in Fig.~\ref{doca}. 
We employ RL algorithms to train the scheduler. In the next subsection, we introduce the concept of RL, followed by describing how we apply it in our context.

\begin{figure}[!t]
	\centering
	\includegraphics[width=\columnwidth]{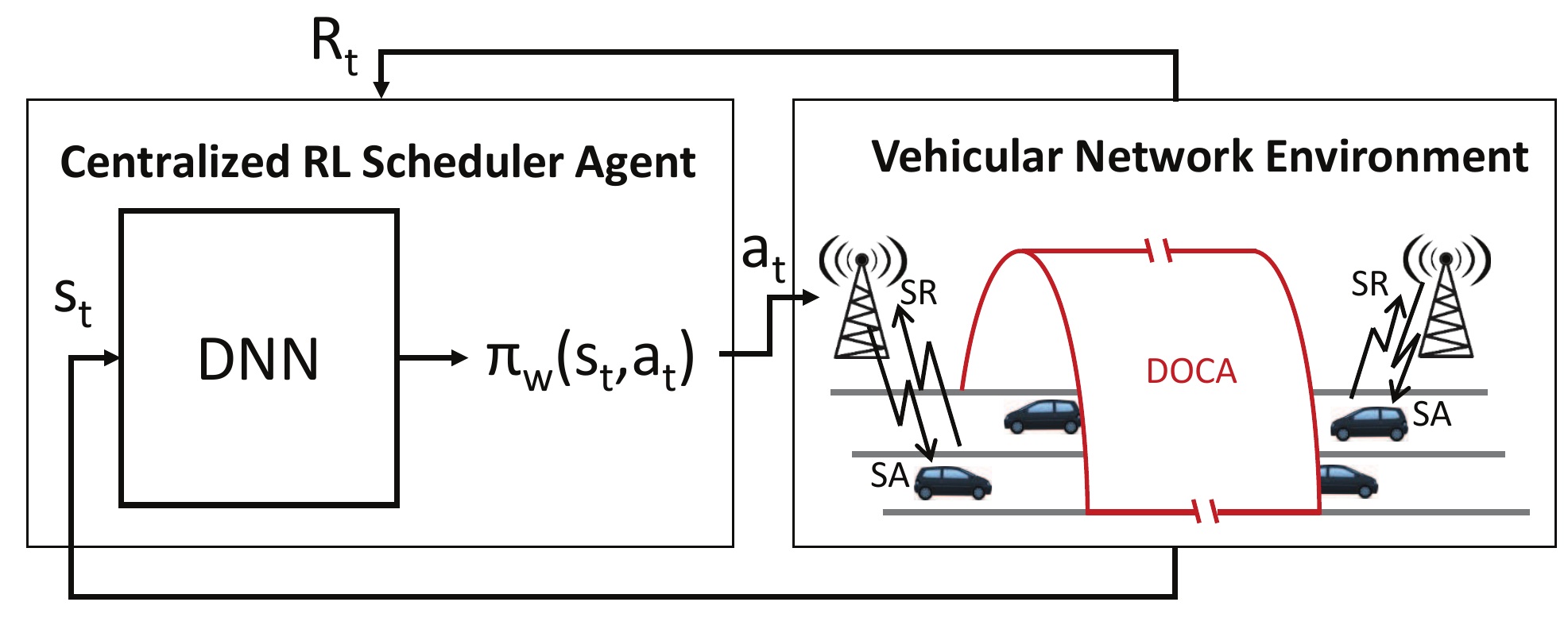}
	\caption{MDP framework applied to our scheduling problem.}
	\label{RL}
\end{figure}	

\begin{figure}[!t]
	\centering
	\includegraphics[height=4cm]{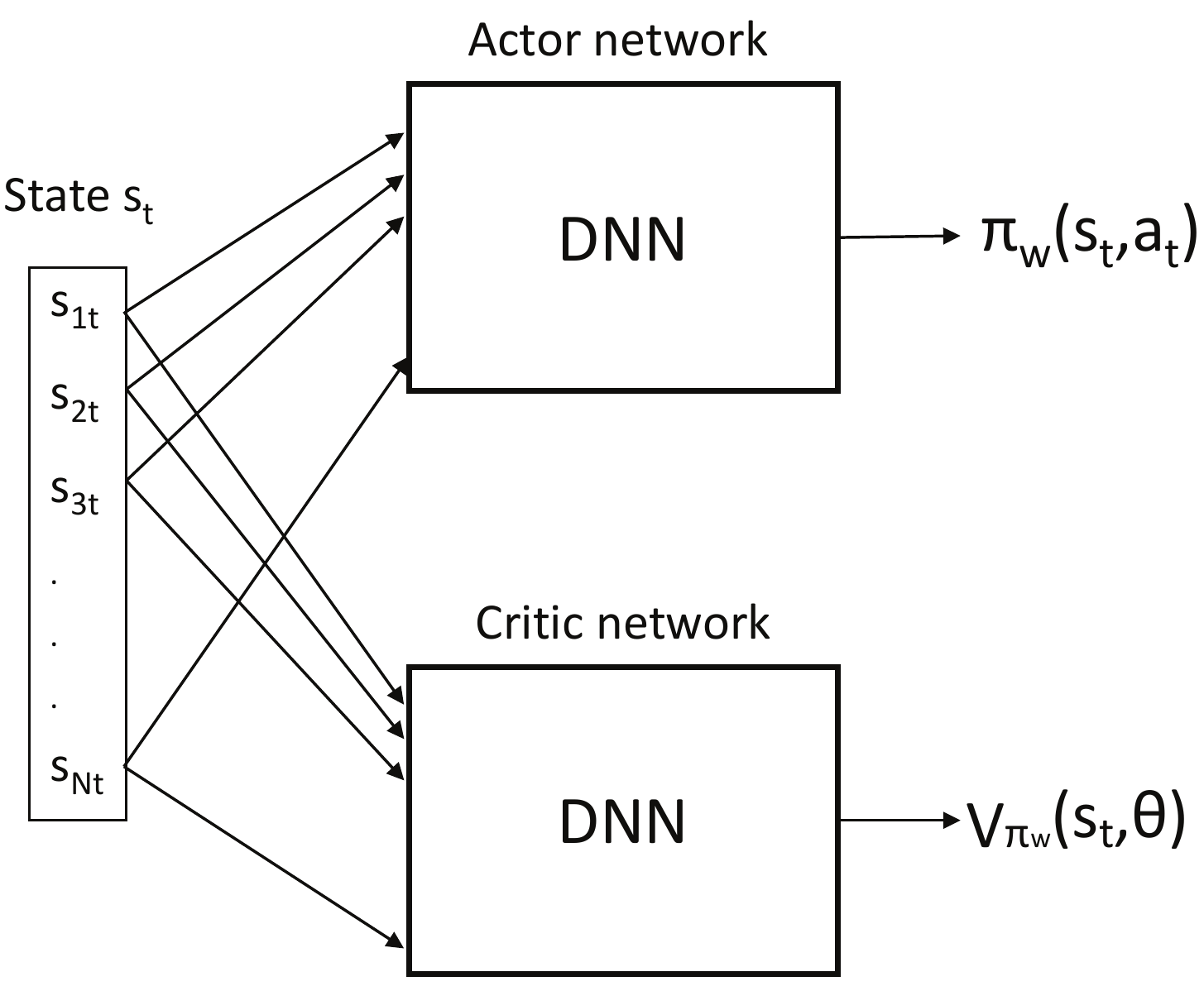}
	\caption{Components of the A3C algorithm.}
	\label{A3C}
\end{figure} 

\subsection{Reinforcement Learning}
\label{Reinforcement}

We apply RL 
to determine the scheduling policy. RL considers a setting where an agent is interacting with its environment by applying a policy which determines the agent's behavior, or action, based on the available information, or perceived states of the environment. By applying an action, the agent receives a reward signal and new state information from the environment. The agent's objective is to maximize total reward collected in the long run. 

The formal framework for RL is defined in the context of Markov decision processes (MDPs). In an MDP, at each time step $t$, the agent observes the state of the environment $S_t$, applies an action $A_t$ following a policy $\pi(s_t, a_t)$, and observes a reward signal $R_{t+1}$ from the environment. (Note that the policy is a function of time; however, for the sake of simplicity, we remove time from its notation.). After applying the action, the state of the environment transitions to a new state $S_{t+1}$. The goal of the agent is to maximize the expected reward $\lim_{t\rightarrow\infty}\mathop{\mathbb{E}}_t[R_t | A_{0:t-1}\sim\pi]$, where $A_0$, $\cdots$, $A_{t-1}$ are the previous actions taken according to the policy $\pi$. Note that we do not use discounting in the definition of the expected reward~\cite{sutton1998reinforcement}. Figure~\ref{RL} depicts how this framework can be applied to our scheduling problem. 
Whenever a new vehicle enters the DOCA, a new action should be taken by the agent. The action consists of assigning a TB to the vehicle. The assignment is performed according to a policy $\pi : \pi(s_t, a_t) \rightarrow[0,1]$, which defines a probability distribution over the set of available actions. $\pi(s_t, a_t)$ is the probability that action $A_t=a_t$ is taken in state $S_t=s_t$.

\subsection{Training Algorithm}
Given the possible number of resources and vehicles (both possibly in thousands or more), 
there are many potential pairs of $(state,action)$, makes tabular solutions infeasible for this problem~
\cite{sutton1998reinforcement}. We therefore propose to apply approximate solutions, where the policy is represented by a deep neural network (DNN) with a set of adjustable policy parameters $w$, i.e. $\pi_w(s_t, a_t)$. The benefits of applying such solution are twofold: i) it makes the learning process much faster, as the number of policy parameters are typically much smaller than the number of $(state,action)$ pairs; and ii) it learns through raw observations and requires no prior information about the task in hand, and the model of the environment. To train the policy parameters, we use A3C \cite{Mnih2016} which applies an actor-critic method.    

The actor-critic algorithm used in our solution involves training two DNNs, one which is used to represent the policy, referred to as the actor network, and the other one which is used to represent state values, referred to as the critic network (see Fig.\ref{A3C}). The value of a state under the policy $\pi_w$ is defined as the expected rewards received by that state in a long run. We denote by $V_{\pi_w}(s_t, \theta)$ the value of state $s_t$  while following $\pi_w$, represented by a critic network with value parameters $\theta$. These state values are used as a critic when training the policy parameters. Similar to \cite{Mnih2016}, we apply policy gradient methods to train the parameters of the actor and the critic networks, i.e. $w$ and $\theta$. Thanks to the policy gradient theorem \cite{sutton1998reinforcement}, an exact expression on how the performance is affected by the policy parameters can be driven for such methods. This ensures performance improvement at each step and hence provides strong convergence properties for policy gradient methods. Besides, using separate networks to represent the state values and the policy removes the possible bias and dependencies introduced when applying policy gradient methods, which in turn accelerates the learning. A detailed discussion on how the training is performed can be found in \cite{sutton1998reinforcement}.   

\begin{figure}[!t]
	\centering
	\subfigure[Radio resource pool with its occupancy status represented with colors]{\label{Pool}\includegraphics[width=\columnwidth]{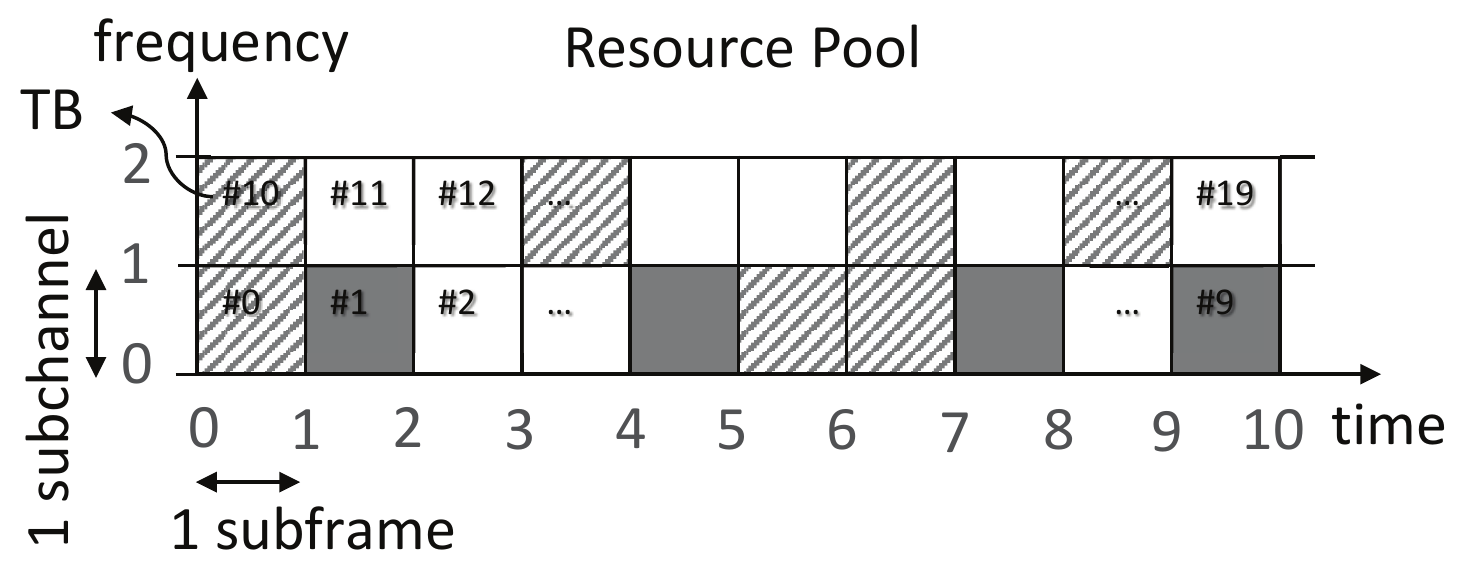}}
	\subfigure[State information used in E1]{\label{State1}\includegraphics[width=\columnwidth]{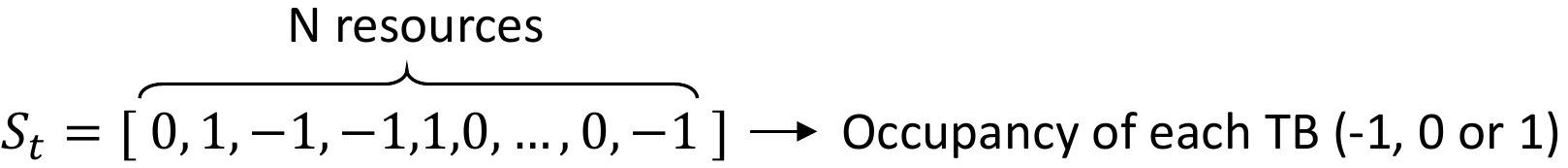}}
	\subfigure[State information used in E2]{\label{State2}\includegraphics[width=\columnwidth]{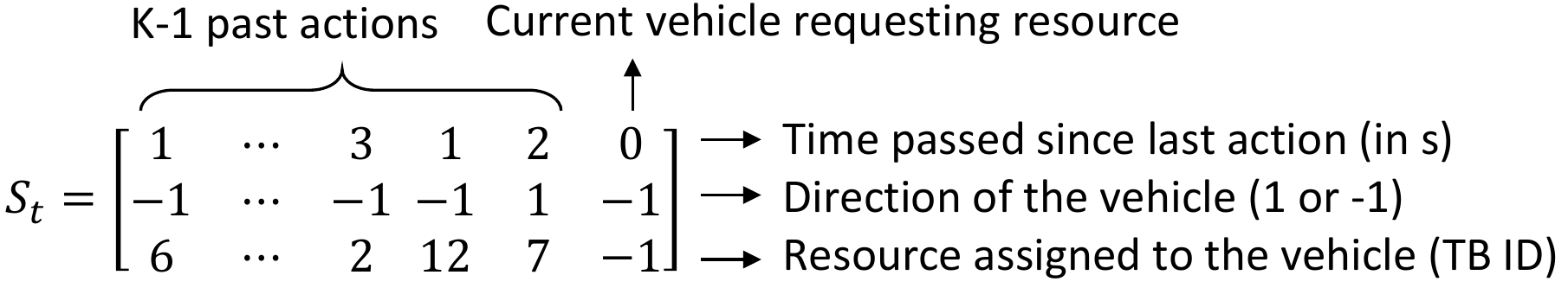}}
	\caption{Example representations of different state information.}
	\label{States}
\end{figure}

\subsection{Implementation}  
\label{Implementation}    
We have implemented different structures of DNNs for E1 and E2, utilizing different definitions of state and reward, as described below.

The implemented DNN (both actor and critic) for E1 consists of 2 convolutional layers followed by 2 fully connected ones. All layers have ``tanh'' as the activation function \cite{nature}, except the last one using linear function in the case of critic network, and ``softmax'' function in the case of actor network, to output the value function, and the action probabilities respectively.


In E1, state information input to the DNNs represents the number of vehicles each resource is assigned to, and the output of the actor network provides the policy determining which resources are to be assigned to the vehicle entering DOCA. Therefore, both state information and policy have the same size $N$ equal to the number of total resources. An example of state representation is provided in Fig. \ref{Pool}. A resource pool consisting of 2 subchannels and 10 subframes (i.e., containing a total of 20 TBs with identifiers (IDs) \#0-19) is utilized by the scheduler. We represent the resource occupancy of the pool with colors. White colored TBs (e.g., TB \#2 and \#11) indicate that they are not assigned to any vehicles inside DOCA, striped ones (e.g., TB \#0 and \#10) indicate that a TB is assigned to a single vehicle, and dark-gray-colored TBs (e.g., TB \#1 and \#4) are assigned to more than a single vehicle. This way, we quantize the number of vehicles each TB is assigned to, which reduces the state-space considerably and consequently accelerates the learning process. The quantized state information is still sufficient as the resources used by any number of vehicles greater than one vehicle will result in collisions, due to the assumption that all vehicles are in the transmission range of each other in E1. 

We therefore represent each element of the state vector with $-1$ if a resource is not scheduled to any vehicle, with $0$ if it is scheduled to a single vehicle, and with $1$ if that TB is scheduled to more than one vehicle inside DOCA. Accordingly, the state vector of the example in Fig.~\ref{Pool} is $S_t = [0,1,-1,-1,1,0,...,0,-1]$, as provided in Fig.~\ref{State1}. Given $S_t$, one expected action would be $A_t=[\#2]$, namely the scheduler assigning the TB \#2 to a new vehicle entering DOCA, which will not result in any collisions and half-duplex errors with the vehicles already traversing the DOCA.

The goal of learning is set to maximize the reliability of transmissions taking place in DOCA. We use the reliability metric PRR from the 3GPP standards~\cite{3gppTR36885} confined to DOCA. Namely, for a single transmitted packet, we measure the number of vehicles successfully receiving the packet divided by the total number of vehicles within the DOCA, and average it over all transmissions taking place in DOCA. Specifically, after each action, the reward collected from the environment is defined to be $+10$ in case $PRR\geq{90\%}$ for all transmissions, and $-10\times(1-min(PRR))$, otherwise, where minimum PRR of any of the transmissions is used.  

In E2, we consider a more realistic and complex environment, where the reuse of the radio resources is possible. For such decisions to be given by the scheduler, a state information containing quantized counts of resource occupancy as in E1 will not be sufficient. It is critical to know which resource was scheduled, to which and how many vehicles, and also when it was scheduled, as vehicles travel across the DOCA using the same resource. Therefore, we utilize the following structure. The state information of E2 has a matrix structure of size 3 by $K$, as illustrated in Fig.~\ref{State2}, where each of the first $K-1$ columns contains information corresponding to each of the $K-1$ previous actions taken, and the last column representing the information about the current vehicle requesting resource from the scheduler, just before entering the DOCA. For each action or column, the first row represents the time passed since the previous action was taken, rounded to the closest integer number of seconds (e.g., $0$, $1$, $2$, etc.). The second row is the direction of the vehicle for which the action is taken (i.e., $1$ for from west to east, and $-1$ for from east to west). Finally, the third row is the TB ID scheduled by that action (e.g., $7$, $12$, etc.). For the last column, i.e., the current vehicle requesting resource, 0 is put in the first row as the time passed, it's direction is entered in the second row, and a dummy variable $-1$ is inserted into the third row, as it's resource is to be assigned by the current action to be taken. Followingly, in the next state, the elements of the matrix will be shifted left by one, with the entries of the last vehicle being updated with actual values, and the information of the next vehicle entering the DOCA that needs to be assigned a resource is appended to the right end of the matrix.

Such a state representation contains all the necessary information from the environment in a compact form, whose size is independent of the number of vehicles and resources available in the network. Differing from E1, rows of the 2D state information from E2 are separately fed into different convolutional layers as the input. Output of each layer are then merged and fed to the second convolutional layer together, followed by a single fully connected layer. Activation functions of the layers follow the same structure as in E1. In order to avoid under-utilization of any resources by the scheduler, we have also modified the reward definition in E1 as $-10\times(1-min(PRR)) - N_{0}$, where $N_{0}$ is the number of resources that are not assigned to any vehicle in that state. 

We utilize the A3C algorithm with multiple agents, each interacting with a different instance of the environment, i.e., with different random seeds of the simulation. Each simulation starts with a random assignment of resources to the vehicles, and a random action taken. After a certain period of interaction and experience with the environment, called an epoch, each agent reports the collected state-action-reward sequences to the central coordinator, which in turn updates the parameters of the DNNs used for learning the policy and the state values.

%% file: Results.tex
\section{Evaluation}
\label{Results}

\subsection{Simulation Setup}

As shown in Fig.~\ref{doca}, we assume a cellular network system with a DOCA. Inside the DOCA, the vehicles travel with a constant speed across a straight section of a highway. 
Vehicles have an average inter-vehicle spacing of $2.5$~s times their speed, following a spatial Poisson distribution at each lane. For simulations, we use a DOCA of $500$~m in length with a single lane per direction, each $4$~m wide. The number of vehicles residing in DOCA is assumed to be constant over time due to the stationary distribution of vehicles among the lanes and the directions.

Vehicles are assumed to generate CAMs of size $190$~B, each occupying a single TB, with a periodicity of $100$~ms. Transmissions take place inside the resource pool utilized by the scheduler. As indicated in Fig. \ref{Pool}, a resource pool consists of set of time and frequency resources, i.e., subframes and subchannels, hence repeating over time with the periodicity equal to the number of subframes of the pool, which we also call as ``scheduling'' or ``control period''. Whenever a vehicle generates a message, it transmits it using the resource assigned by the scheduler, within the next available control period. Accordingly, the time-length of the pool also bounds the maximum latency that a transmission may experience. Initial CAM generation is randomized across the subframes of the first simulated control period. Therefore, the periodically repeated transmissions always take place within the same control period, hence confined to the defined resource pool.

For E1, we consider three scenarios, designated E1-A, E1-B, and E1-C, which differ with respect to vehicle densities and the amount of resources. In scenario E1-A, $10$ vehicles reside in DOCA, all traveling at $140$~km/h, where a resource pool consisting of $1$ subchannel and $10$~subframes is utilized by the scheduler. In scenario E1-B, $12$ vehicles travel at $140$~km/h, this time having a resource pool of $2$ subchannels, and $10$ subframes. In the latest scenario, E1-C, there are $24$ vehicles traveling at $70$~km/h residing in DOCA, utilizing a resource pool with the same size of $2$ subchannels, and $10$ subframes.
Our choice of the scenarios is motivated by the goal of representing the following three cases of network condition;
E1-A: loaded, without half-duplex (HD) constraint, E1-B: under-loaded, with HD constraint; and E1-C: over-loaded, with HD constraint. 

In E2, we consider a single scenario where $30$ vehicles are traveling at the speed of $50$~km/h across the DOCA, and where a resource pool of $2$ subchannels by $10$ subframes is available.
This scenario is used to evaluate the potential of our RL solution on reusing resources which will overcome the drawbacks of the overloaded situation. In order to enable resource reuse within the considered DOCA of $500$~m size, transmission powers of the vehicles are reduced to $-5$~dBm (as opposed to transmitting with the maximum power of $23$~dBm in E1). This way, the power received beyond $100$ m away from the transmitter is reduced to around noise power level, which in turn enables reusing the same resource at around a distance of $200$ m. Moreover, the channel model in~\cite{3gppTR36885} for the ``freeway case'' is assumed between each vehicle in the environment, whose details are appended to Table \ref{tableScenario}. Due to these assumptions, PRR in E2 is measured for the receivers at up to a distance of 100 m.
	
The complexity of E2 results in longer simulation times, mostly due to computing SINRs upon each reception so as to determine if a packet is successfully received or not. In order to reduce the training time, the agents are initially trained in an environment simplified in terms of the propagation model, where the received power is assumed to be constant up to $120$ m, and $-\infty $ afterwards, i.e., a transmission-range based interference model.

The evaluated environments and scenarios are summarized in Table \ref{tableScenario}, together with the corresponding values of the utilized parameters in each of them.

\begin{table}[!t]
	\renewcommand{\arraystretch}{1.1} 
	\caption{Simulation Parameters}
	\label{tableScenario}
	\resizebox{\columnwidth}{!}{%
		\centering
		\begin{tabular}{|l|l|l|l|l|}
			\hline
				& \textbf{E1-A} & \textbf{E1-B} & \textbf{E1-C} & \textbf{E2 }\\
			\hline
			Number of vehicles & $10$ & $12$ & $24$ & $30$\\
			\hline
			Vehicle speed & $140$ km/h & $140$ km/h & $70$ km/h & $50$ km/h\\
			\hline
			\multirow{2}{*}{Resource pool} & $1$ subchannel & \multicolumn{3}{c|}{$2$ subchannels} \\
										   & $10$ subframes & \multicolumn{3}{c|}{$10$ subframes}\\								   
			\hline
			\multirow{2}{*}{DOCA size} & \multicolumn{4}{l|}{500 m of a straight highway,}\\
									& \multicolumn{4}{l|}{1 lane per direction, $4$~m lane width}\\
			\hline
			Vehicle spatial distribution & \multicolumn{4}{l|}{2.5-s distance ahead, with Poisson distribution}\\
			\hline 
			Transmission power & \multicolumn{3}{l|}{$23$ dBm (the maximum value)} & $-5$ dBm\\
			\hline 
			CAM size and periodicity & \multicolumn{4}{l|}{$190$ B, $100$ ms}\\ 
			\hline
			Subframe duration & \multicolumn{4}{l|}{$1$ ms}\\
			\hline
			\multicolumn{5}{|c|}{\textbf{V2V channel model in E2 \cite{3gppTR36885}}}\\
			\hline
			\multirow{2}{*}{Pathloss model} & \multicolumn{4}{l|}{LOS in WINNER+B1 with antenna height = $1.5$ m;}\\
				& \multicolumn{4}{l|}{pathloss at $3$ m is used for distance $<3$ m}\\
			\hline
			\multirow{2}{*}{Shadowing fading} & \multicolumn{4}{l|}{Log-normal distributed with $3$ dB standard deviation,}\\
			& \multicolumn{4}{l|}{and decorrelation distance of $25$ m}\\
			\hline	
		\end{tabular}
	}
\end{table}

\begin{figure*}[!t]
	\begin{center}
		\subfigure[Scenario E1-A.]{\label{Res1}\includegraphics[width=0.5\columnwidth]{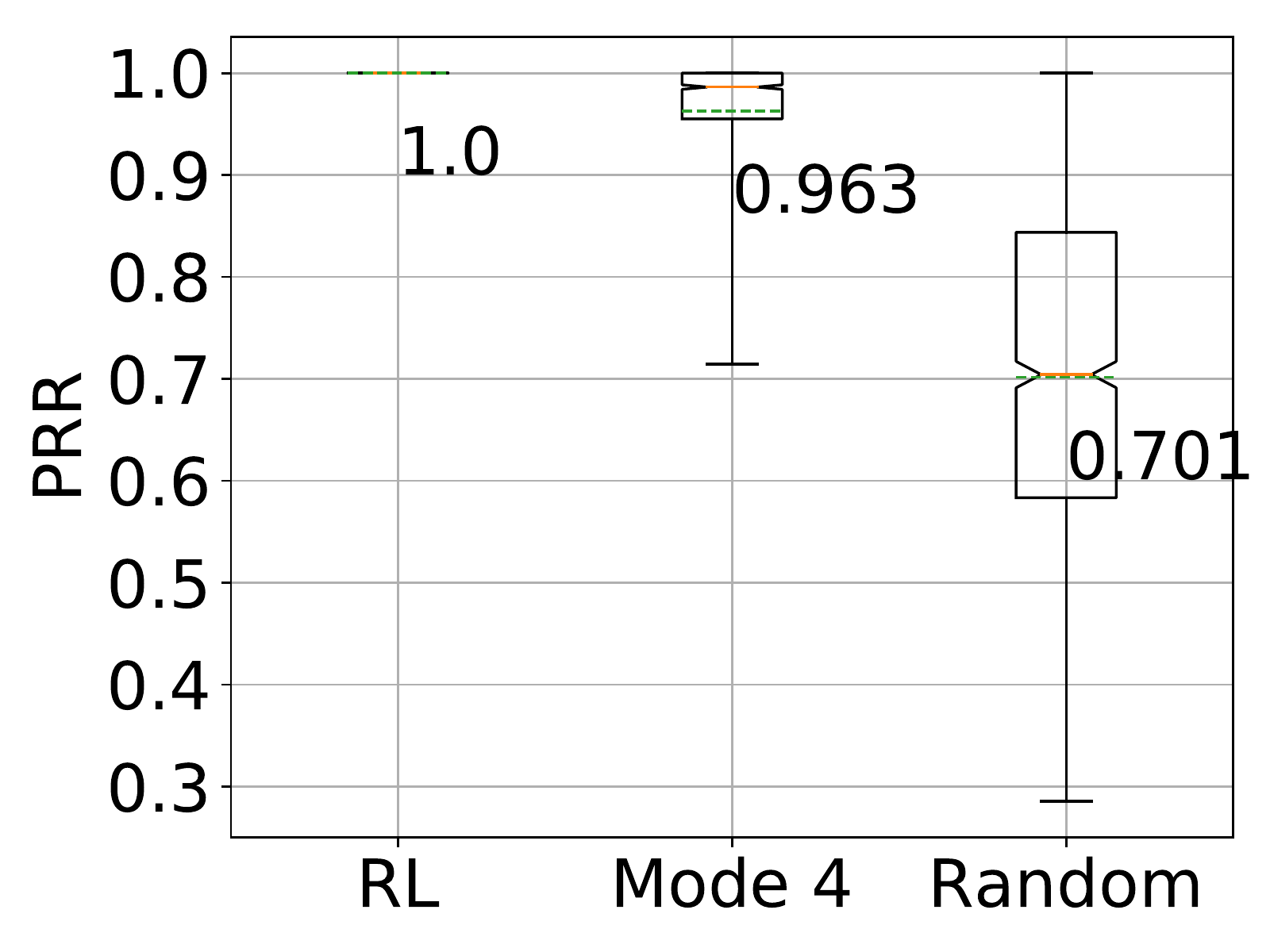}}
		\subfigure[Scenario E1-B.]{\label{Res2}\includegraphics[width=0.5\columnwidth]{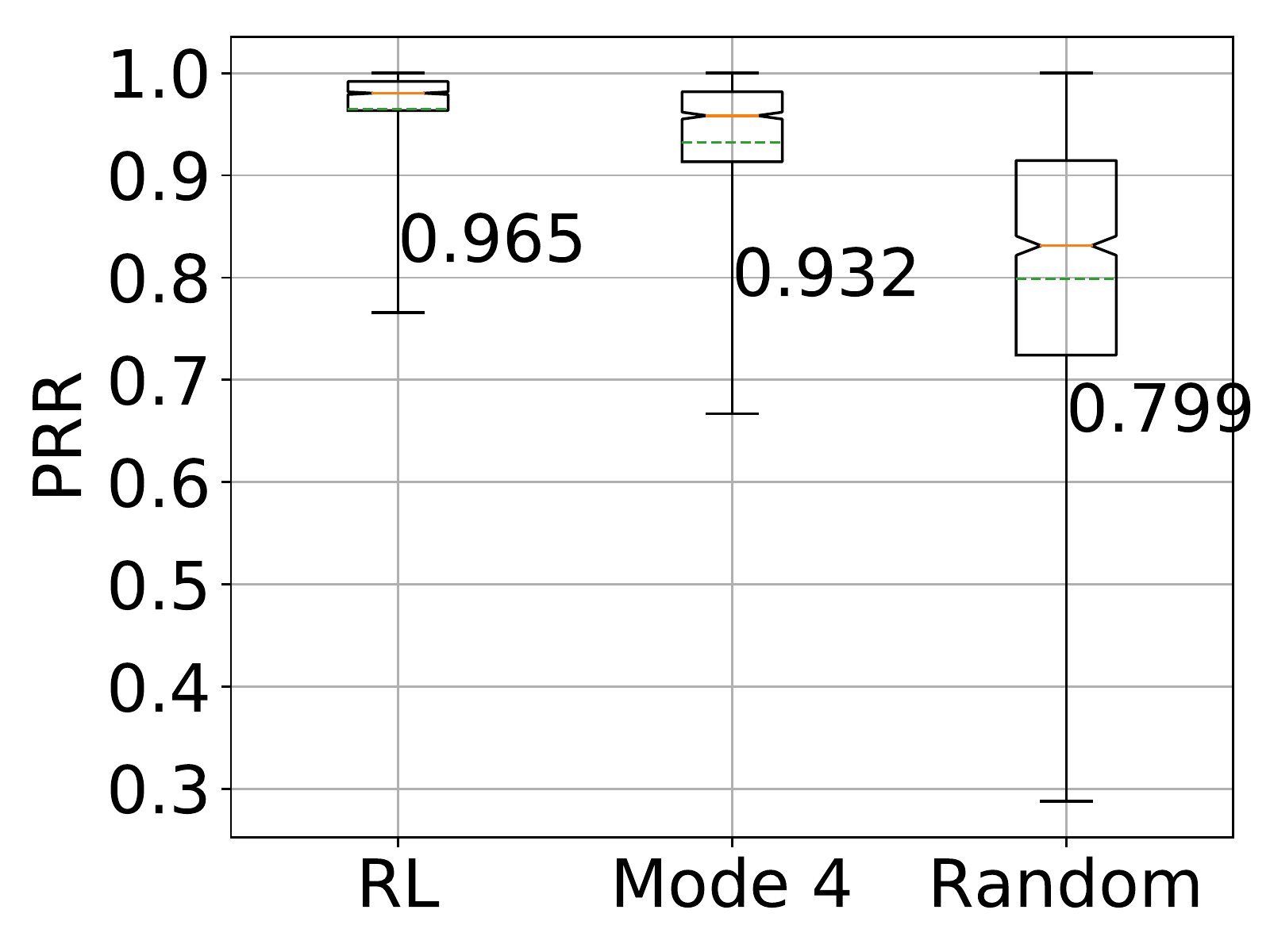}}
		\subfigure[Scenario E1-C.]{\label{Res3}\includegraphics[width=0.5\columnwidth]{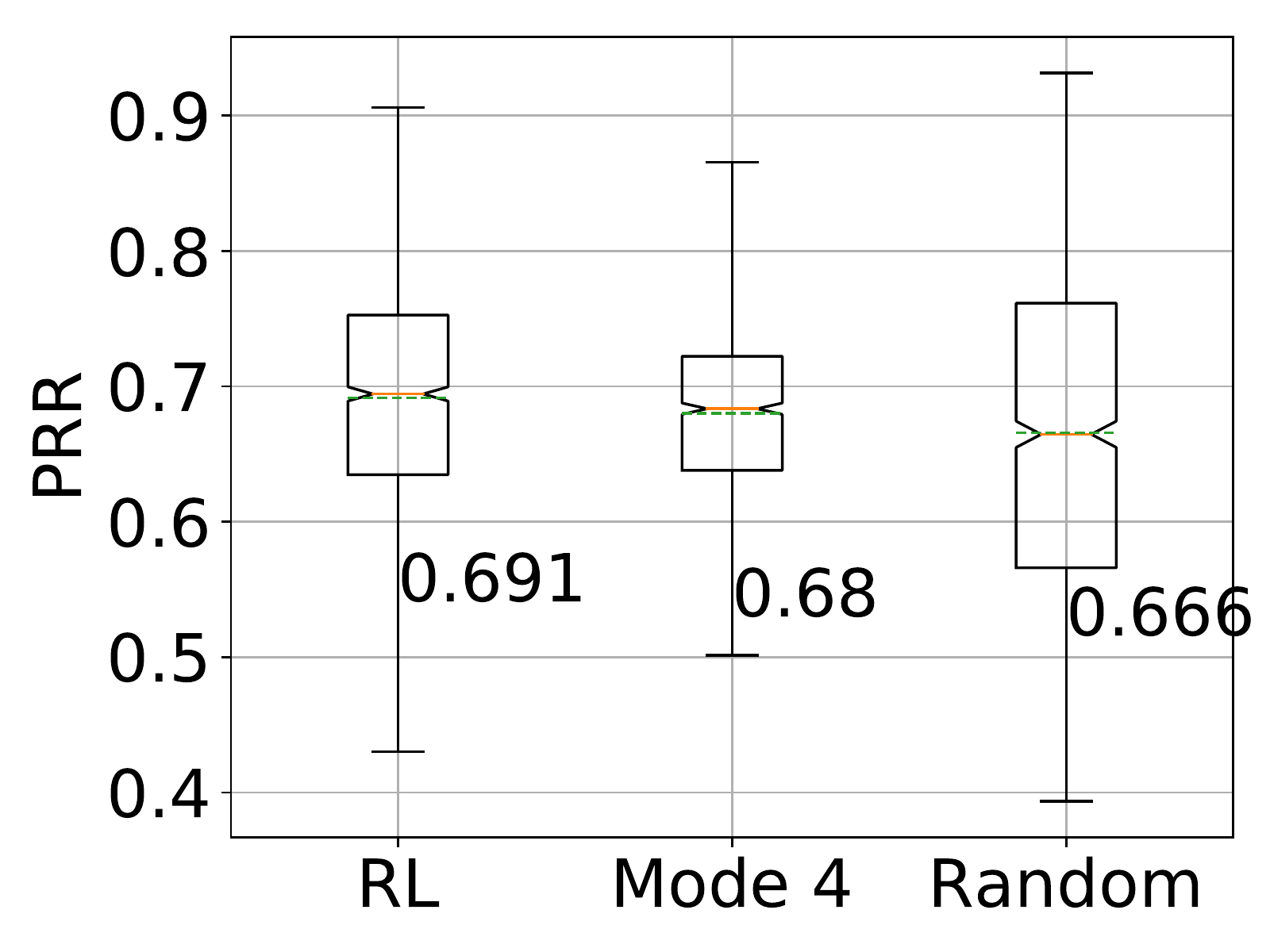}}
		\subfigure[Scenario E2.]{\label{Res4}\includegraphics[width=0.5\columnwidth]{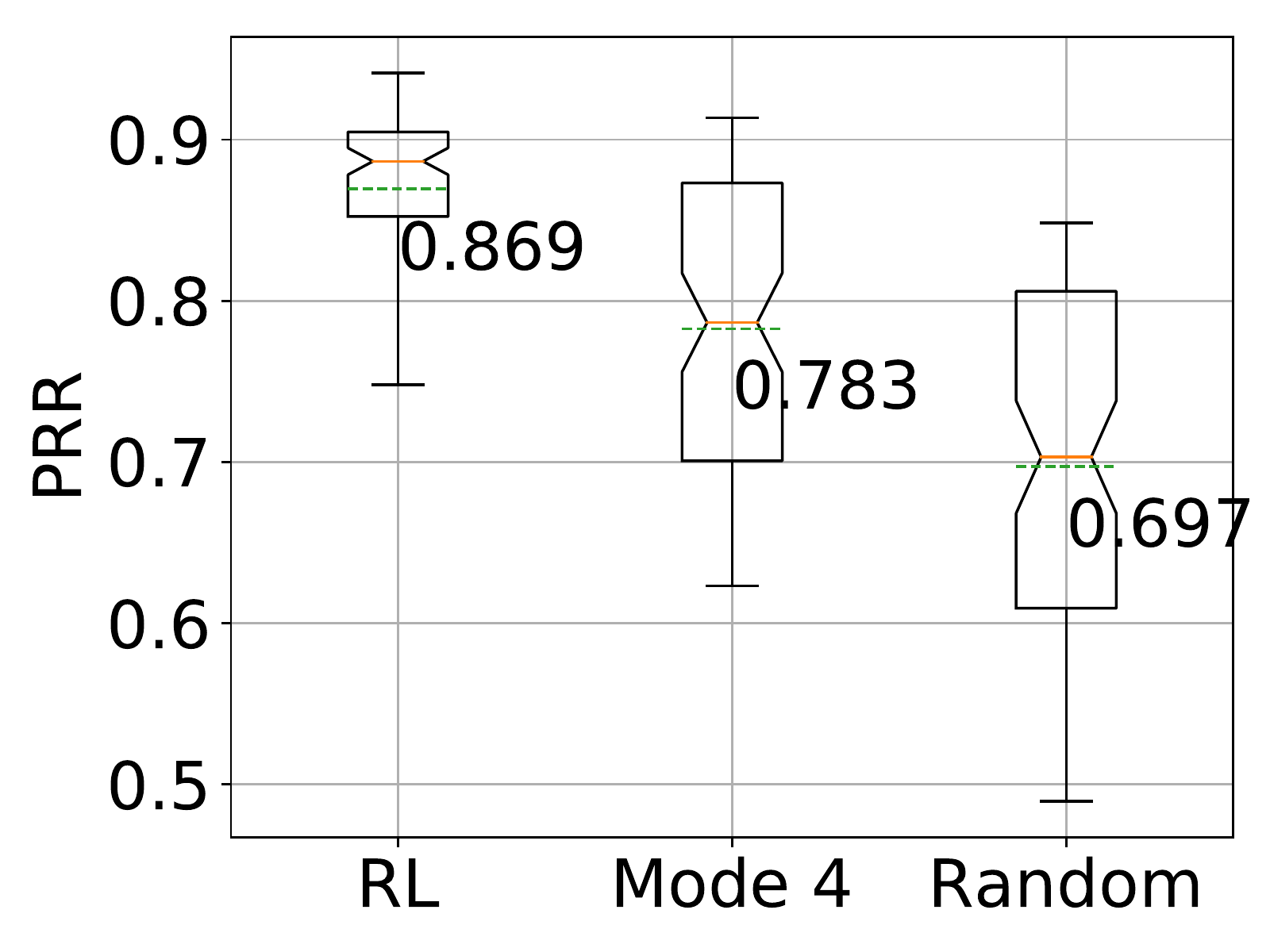}}
	\end{center}
	\caption{Median (red), mean (green, dashed), 25$^{th}$ and 75$^{th}$ percentile (box), and 1$^{st}$ and 99$^{th}$ percentile (whiskers) of PRR for the proposed centralized RL scheduler, distributed Mode-4 algorithm~\cite{3gppTR36885}, and a centralized scheduler assigning random resources.}
	\label{fig:Results}
\end{figure*}

\subsection{Comparison}
\label{Comparison}

In this section, we evaluate and compare the performance of the trained centralized RL scheduler with two baselines: the sensing-based distributed scheduling Mode-4 from the 3GPP standard~\cite{3gppTR36885}, and a centralized scheduler assigning random resources to the vehicles entering DOCA. The key performance indicator (KPI) we are interested in is the average PRR. Simulation of the vehicular environment is carried out using the network simulator ns-3 \cite{ns3}, and the vehicular mobility simulator SUMO \cite{sumoPaper}. Specifically, we have extended an LTE-D2D module for ns-3 \cite{D2D} to support LTE-V functionalities, including Mode-4. We report the results in Fig. \ref{fig:Results}. The results shown in the figure were 
collected over more than $1000$ resource assignments/actions.



\begin{table}
	\renewcommand{\arraystretch}{1.1} 
	\caption{Training Parameters}
	\label{tableTraining}
	\resizebox{\columnwidth}{!}{%
		\centering
		\begin{tabular}{|l|l|l|l|l|}
			\hline
			& \textbf{E1-A} & \textbf{E1-B} & \textbf{E1-C}& \textbf{E2}\\ 
			\hline
			State information size & $10\times1$ & $20\times1$ & $20\times1$ & $(K=30)\times3$\\
			\hline
			Number of actions per epoch & $20$ & $30$ & $48$ & $120$\\
			\hline
			Number of training epochs & $400$ & $1400$ & $1200$ & $930$\\
			\hline
			\multirow{2}{*}{Learning rates of actor-critic} & \multirow{2}{*}{$10^{-4}$} & \multirow{2}{*}{$10^{-4}$}  & $10^{-4}$, and $10^{-5}$& \multirow{2}{*}{$\frac{10^{-3}}{ \left \lfloor{1 + 0.01\times \#ep^{1.1}}\right \rfloor}$}\\
			&  			& 	 		&   for $\#ep>1000$&\\ 
			\hline
			Layers of actor-critic DNNs & \multicolumn{3}{l|}{2 convolutional + 2 fully connected} & 2 conv. + 1 FC\\
			\hline
			Number of agents & \multicolumn{4}{l|}{$16$}\\
			\hline
		\end{tabular}
	}
\end{table}

For E1-A, RL scheduler is able to perform at $100\%$ PRR (after eliminating the transient phase that starts from the state of randomly assigned resources), which is achieved by learning to allocate time-orthogonal resources to each vehicle in DOCA. As the number of vehicles inside the DOCA is equal to the number of resources, no collision would occur, and all the vehicles can hear each other all the time. 
Mode-4 is able to achieve a mean PRR of $96.3\%$, where the performance degradation comes from the randomness in its resource selection algorithm. After each sensing period, vehicles select the resource to transmit randomly among the best\footnote{``Best'' in this context is defined as the lowest energy sensed on the resource.} $20\%$ resources according to their sensing results (for details, see~\cite{3GPPTS36300}). In our case, each vehicle selects one of the two best resources out of 10 at random, which results in collisions if an occupied resource is selected. As one of the two selected resources will always be occupied for the case of the last (10$^{th}$) vehicle selecting a resource, collision happens with a probability of 5$\%$ ($1/2\times1/10$) on average, which is in line with our simulation results. The scheduler assigning random resources acts as a reference for the remaining two algorithms, as it performs the worst with a mean PRR of $70.1\%$. The optimal performance, however, could be also achieved using a round-robin scheduler assigning time-orthogonal resources to the vehicles entering DOCA. In that sense, scenario E1-A serves as a sanity-check, where RL scheduler performs optimally. 

For E1-B, performance of both RL scheduler and Mode-4 is degraded, due to introduced HD constraint in the environment. Whenever a vehicle transmits, it does not hear the other transmissions taking place at the same subframe on the next subchannel. Nevertheless, RL can achieve a performance of $96.5\%$ average PRR as compared to Mode-4 ($93.2\%$ average PRR). The strategy that the RL scheduler learns in this scenario is to allocate resources orthogonal both in time and frequency as much as possible. As there are 2 more vehicles than the number of subframes, RL scheduler tries to assign them to different subchannels, rather than assigning to the occupied subchannel at each subframe, hence most of the time resulting only in half-duplex reception errors among 2 vehicles instead of any collision error affecting the reception of all vehicles. Hence, again, the RL scheduler manages to find the near-optimal solution. On the other hand, random resource allocation performs better than in Scenario A, as the network is in an under-loaded condition with higher probability of assigning non-colliding resources, compared to a loaded one.

Scenario E1-C represents the overloaded network conditions, in addition to the HD constraint. Therefore, collisions are unavoidable in any case as all vehicles are assigned resources (i.e., no admission control), which results in a considerable amount of performance degradation in case of all algorithms. In this scenario, RL scheduler develops a strategy where it tries to maximize the number of non-colliding resources, namely assigning them orthogonally in time and frequency as much as possible, as in E1-B, this time scheduling all the remaining vehicles onto one or two resources where they collide. In the best case, 19 vehicles in DOCA are scheduled to orthogonal resources, and the remaining ones are all being assigned the single resource left, which results in a mean PRR of about $75\%$. RL scheduler performs slightly better than Mode-4, and provides a mean PRR of $69.1\%$.

Scenario E2 allows for the reuse of the resources, which results in an overall better performance compared to the overloaded case of scenario E1-C. PRRs up to $94\%$ are achievable by RL, even in the case of higher number of vehicles in DOCA. Compared to E1, RL makes use of additional state information provided by the environment, as explained in Section \ref{Implementation}. Looking at specific state-action pairs, we observe that most of the time RL (re)uses the same resource at either of the directions, with allowing some time gap between each reassignment. Moreover, due to the modified reward definition, it yields very low number of unused resources.

\begin{figure*}[!h]
	\begin{center}
		\subfigure[Scenario E1-A.]{\label{CurveA}\includegraphics[width=0.45\columnwidth]{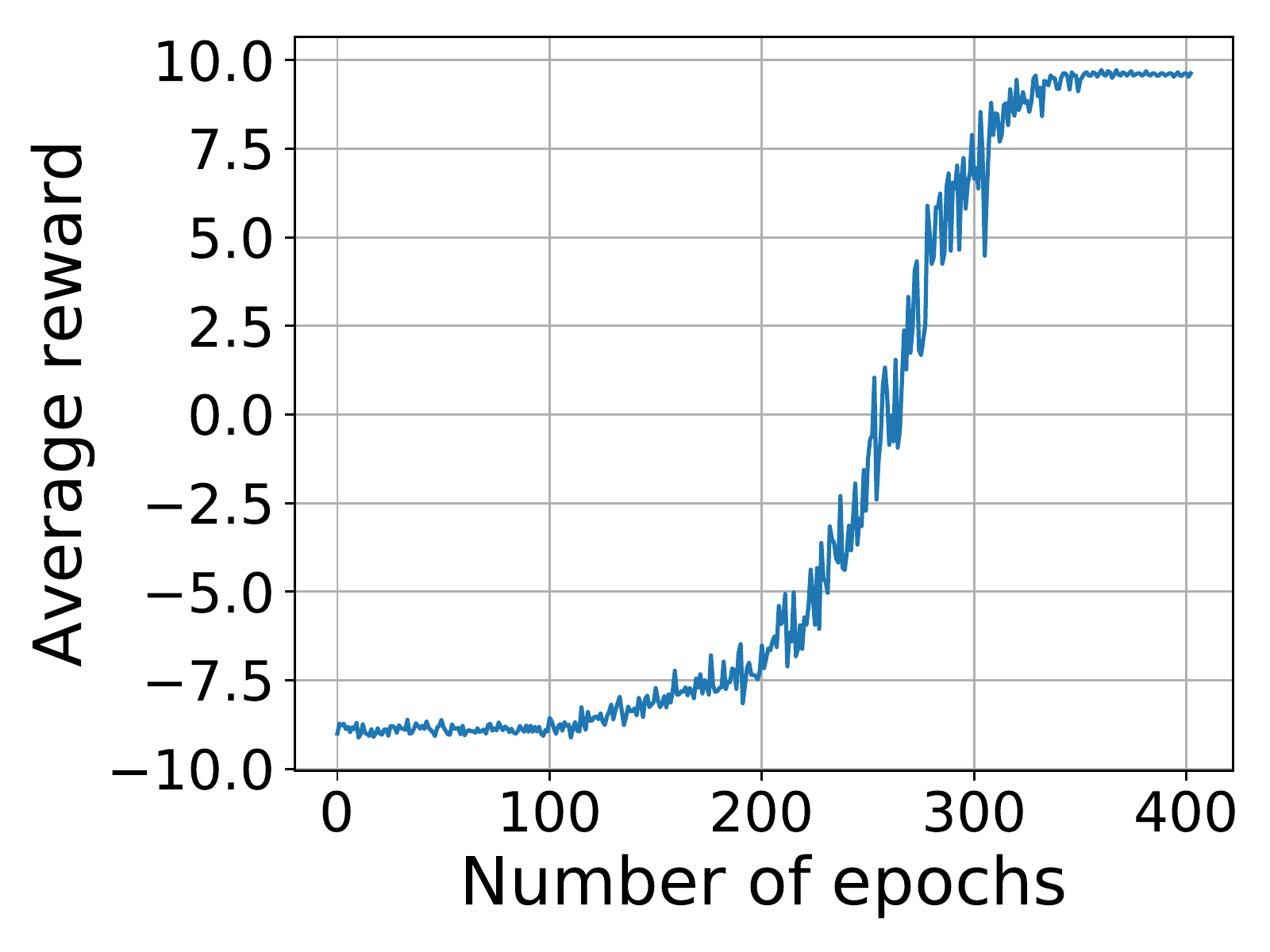}}
		\subfigure[Scenario E1-B.]{\label{CurveB}\includegraphics[width=0.45\columnwidth]{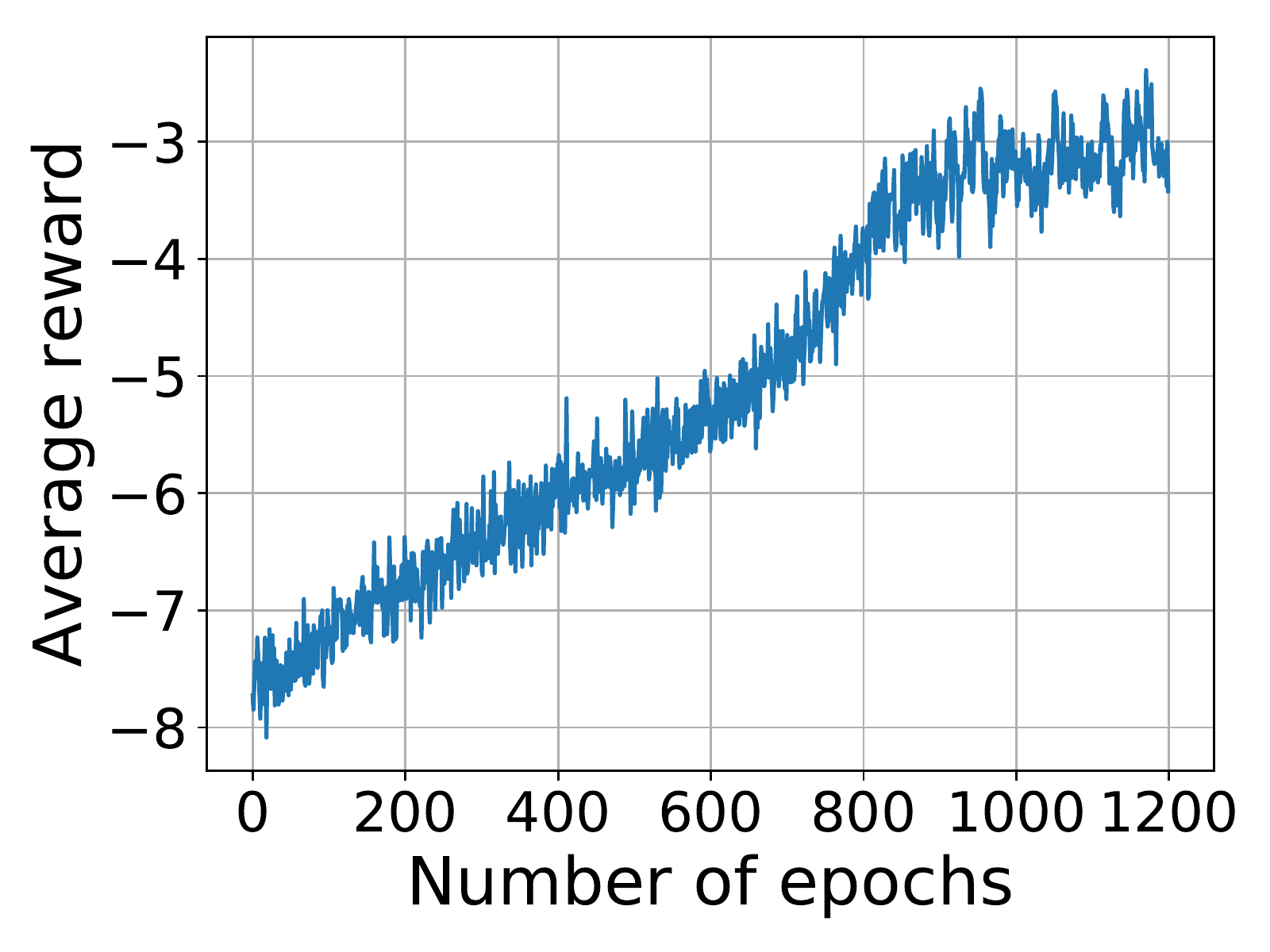}}
		\subfigure[Scenario E1-C.]{\label{CurveC}\includegraphics[width=0.45\columnwidth]{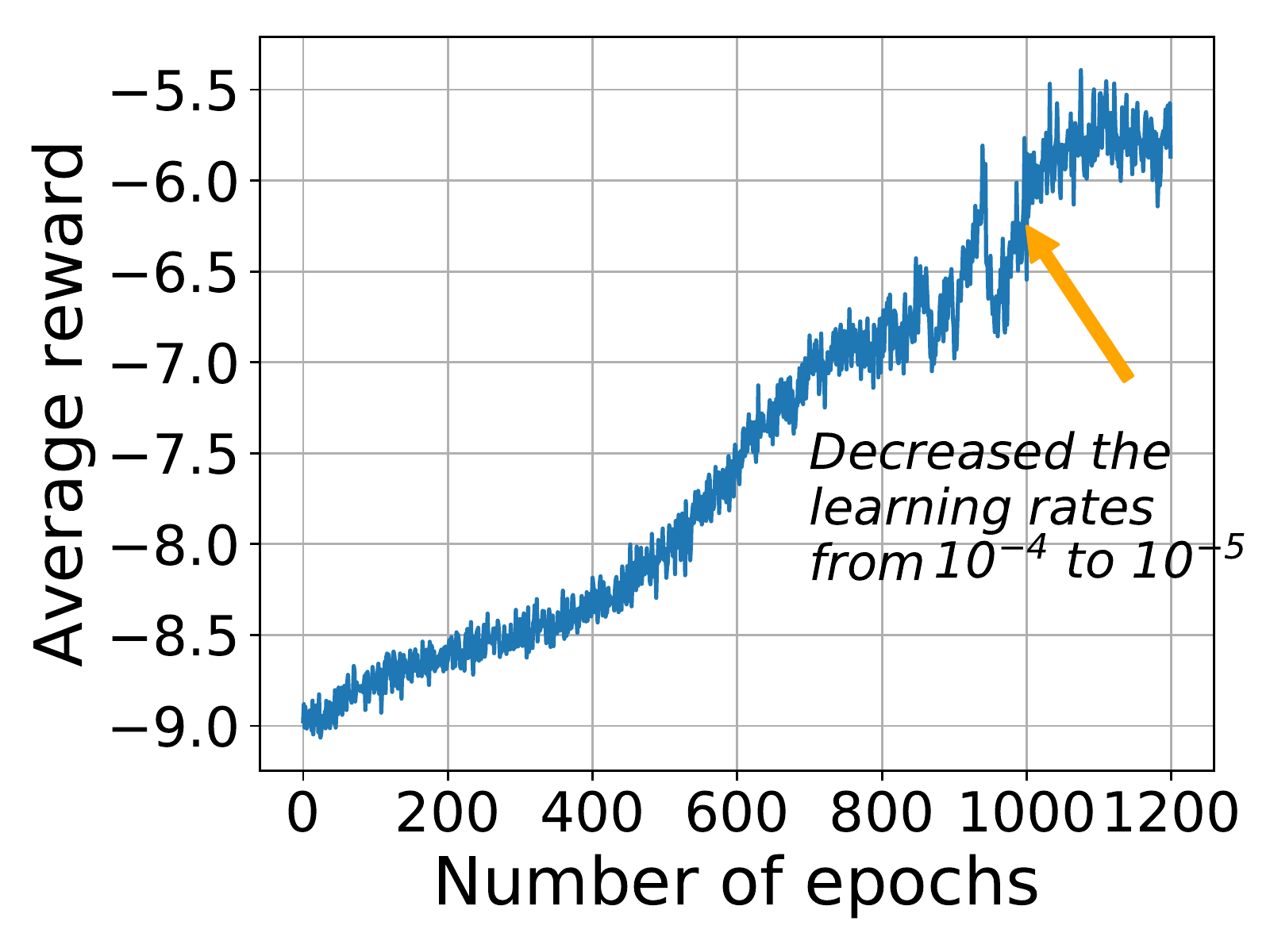}}
		\subfigure[Scenario E2.]{\label{CurveD}\includegraphics[width=0.45\columnwidth]{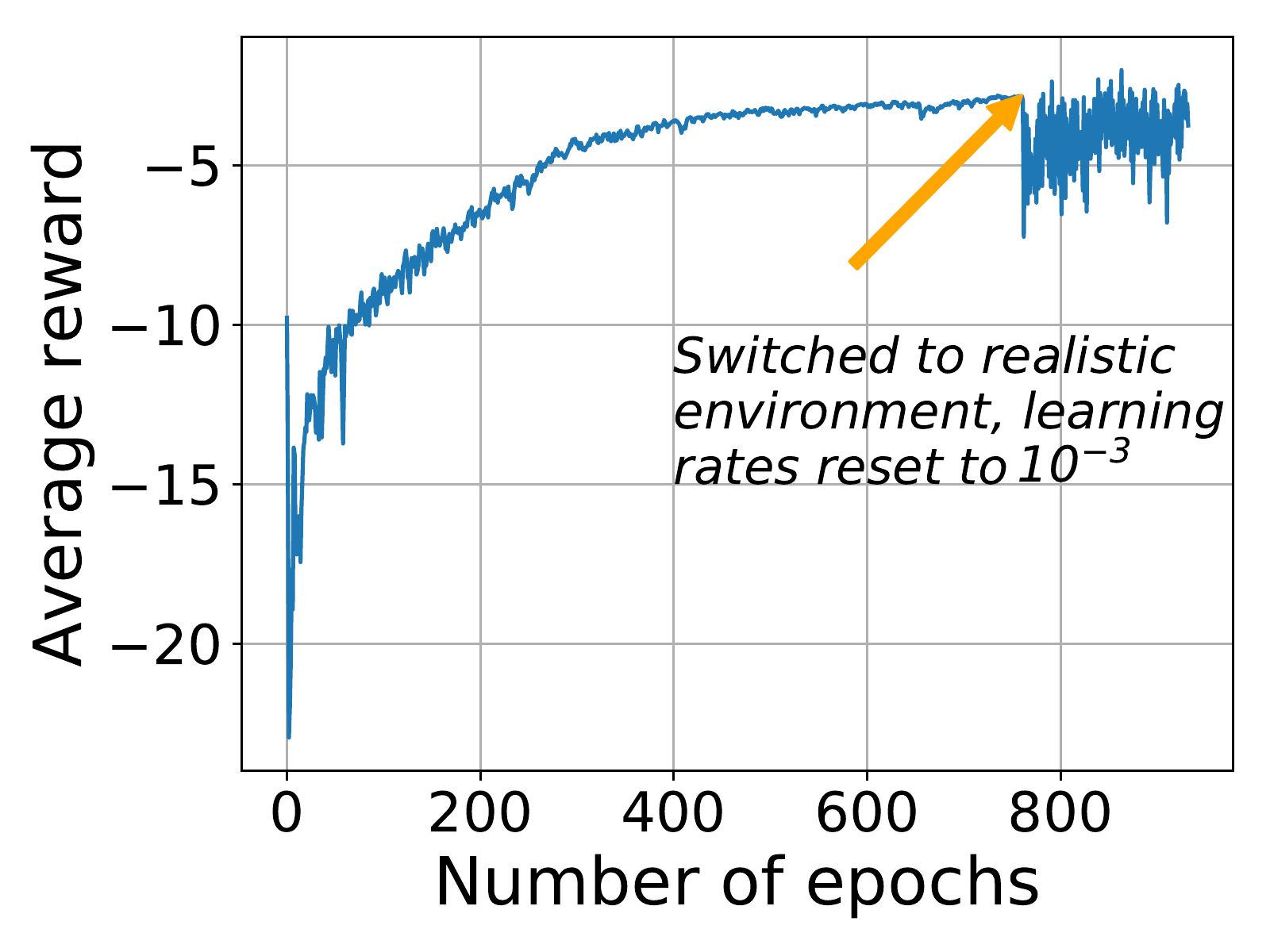}}
	\end{center}
	\caption{Learning curve for each of the environments and scenarios.}
	\label{fig:LearningCurves}
\end{figure*}

\subsection{Learning Curves}
To analyze the convergence of the RL scheduler, we show the learning curves in terms of the collected average reward with respect to the number of training epochs in Fig. \ref{fig:LearningCurves} for all four scenarios. Detailed training parameters for each scenario are provided in Table~\ref{tableTraining}. Note that we are more interested in the convergence, rather than the actual value of the average reward that has been converged to.

It takes around $350$ epochs for the algorithm to converge for scenario E1-A to an average reward of around $9.6$. On the other hand, scenarios E1-B, E1-C, and E2 require more epochs for the algorithm to converge to a certain level of average reward, mainly due to larger state-space they contain. Particularly, the algorithm converges to an average reward of around $-3.1$ after around $900$ epochs in E1-B. In E1-C, the algorithm converges to an average reward of around $-5.7$ at around $1200^{th}$ epoch. 
In order to assure convergence for scenario E1-C, we further tuned the learning rates of the actor-critic DNNs, which is shown to have an impact on the learning performance of the A3C algorithm \cite{Mnih2016}. Specifically, we reduced both from $10^{-4}$ to $10^{-5}$ after the $1000^{th}$ epoch, as we started to observe oscillations on the average reward that are also visible from Fig. \ref{CurveC}.
For E2, the agents were able to converge to an average reward of around $-2.7$ on the simple environment they were trained, in around 760 epochs. The agents were then continued to be trained in the actual environment, in also which their performance was evaluated in Section \ref{Comparison}. Due to longer simulation times, number of training epochs in the realistic environment were limited to around 170 epochs. However, it is expected to have better performance with an extended training. Learning rates for each training are reduced exponentially with the number epochs, starting from $10^{-3}$ as in Table~\ref{tableTraining} to enable better converge.

%% file: Conclusions.tex
\section{Conclusions and Future Work}
\label{Conclusion}

We address the problem of resource scheduling for reliable 
V2V communications, where vehicles communicate inside DOCA, an out-of-coverage area delimited by network infrastructure, with the infrastructure connected to a centralized scheduler. 
We implement an RL-based scheduler that learns how to assign resources to the vehicles through interaction with the environment. We studied the performance of our solution through simulations. Since we address the problem of out-of-coverage, we compare our proposed solution with state-of-art distributed resource scheduling algorithms. Our initial evaluations show that the RL scheduler outperforms existing distributed scheduling algorithms, converging to near-optimal solution on a simulated DOCA of small size ($500$~m) with limited number of vehicles and resources.

In the considered environments and scenarios, the centralized scheduler learned to develop strategies that allowed it to: i) assign fully orthogonal resources in scenario E1-A; ii) avoid HD constraint to the extent possible in E1-B; and iii) to group excess transmissions in case of network overload to a small set of resources in E1-C, thus allowing remaining transmissions to have no collisions. Furthermore, in a more realistic environment, E2, it achieved the reuse of resources by taking the direction and arrival time of the vehicles into account, which lead to success in dealing with an overloaded scenario. Moreover, insights from evaluating a simple environment, E1, helped us to better design the RL setting for a realistic environment, E2. Overall, RL scheduler converged to a solution in
few hundred to thousand epochs.

These initial encouraging results 
motivate us for future work, in particular to consider more complex environments. In this paper, we consider vehicles traveling across a DOCA on a straight highway at the same speed, and transmit periodic traffic. Consequently, a static pre-allocation of a single resource is performed by the scheduler. We are further interested in more dynamic environments, where the speeds of vehicles are different and varying with time, over multi-dimensional directions of travel. Accordingly, vehicles can sweep among different resources to better deal with interference from other vehicles in their vicinity. This can be performed by assigning a \textit{vector} of resources to a vehicle, so that it can switch among these resources, or by integrating some aspects of sensing-based solutions such as those described in~\cite{3GPPTS36300}. Furthermore, 
we plan to extend our RL solution into a distributed setting, where BSs located at each end of the DOCA can run an instance of the scheduler, and when the learning is performed in a distributed manner. 

%% file: Acknowledgment.tex
\section{Acknowledgment}
Part of this work has been performed in the framework of the H2020 project 5GCAR co-funded by the EU. The authors would like to acknowledge the contributions of their colleagues from 5GCAR although the views expressed are those of the authors and do not necessarily represent the views of the 5GCAR project.

%% file: main.bbl

%% file: main.bbl
\begin{thebibliography}{10}
\providecommand{\url}[1]{#1}
\csname url@samestyle\endcsname
\providecommand{\newblock}{\relax}
\providecommand{\bibinfo}[2]{#2}
\providecommand{\BIBentrySTDinterwordspacing}{\spaceskip=0pt\relax}
\providecommand{\BIBentryALTinterwordstretchfactor}{4}
\providecommand{\BIBentryALTinterwordspacing}{\spaceskip=\fontdimen2\font plus
\BIBentryALTinterwordstretchfactor\fontdimen3\font minus
  \fontdimen4\font\relax}
\providecommand{\BIBforeignlanguage}[2]{{%
\expandafter\ifx\csname l@#1\endcsname\relax
\typeout{** WARNING: IEEEtran.bst: No hyphenation pattern has been}%
\typeout{** loaded for the language `#1'. Using the pattern for}%
\typeout{** the default language instead.}%
\else
\language=\csname l@#1\endcsname
\fi
#2}}
\providecommand{\BIBdecl}{\relax}
\BIBdecl

\bibitem{3gppTS22186}
{3GPP TR 22.186 V15.0.0}, \emph{{Service requirements for enhanced V2X
  scenarios (Release 15)}}, 3GPP Std., March 2017.

\bibitem{3gppTR36885}
{3GPP TR 36.885 V14.0.0}, \emph{{Technical specification group radio access
  network; Study on LTE-based V2X services}}, 3GPP Std., June 2016.

\bibitem{3GPPTS36300}
{3GPP TS 36.300 V14.3.0}, \emph{{Technical specification group radio access
  network; Overall description; Stage 2}}, 3GPP Std., June 2017.

\bibitem{Gozalvez}
R.~Molina-Masegosa and J.~Gozalvez, ``{LTE-V for sidelink 5G V2X vehicular
  communications: A new 5G technology for short-range vehicle-to-everything
  communications},'' \emph{IEEE Vehicular Technology Magazine}, vol.~12, no.~4,
  pp. 30--39, Dec 2017.

\bibitem{sahin2018radio}
T.~Sahin and M.~Boban, ``{Radio resource allocation for reliable
  out-of-coverage V2V communications},'' in \emph{2018 IEEE 87th Vehicular
  Technology Conference (VTC Spring)}.\hskip 1em plus 0.5em minus 0.4em\relax
  IEEE, 2018, pp. 1--5.

\bibitem{mao2016resource}
H.~Mao, M.~Alizadeh, I.~Menache, and S.~Kandula, ``Resource management with
  deep reinforcement learning,'' in \emph{Proceedings of the 15th ACM Workshop
  on Hot Topics in Networks}.\hskip 1em plus 0.5em minus 0.4em\relax ACM, 2016,
  pp. 50--56.

\bibitem{magazine}
H.~Ye, L.~Liang, G.~Y. Li, J.~Kim, L.~Lu, and M.~Wu, ``Machine learning for
  vehicular networks: Recent advances and application examples,'' \emph{IEEE
  Vehicular Technology Magazine}, vol.~13, no.~2, pp. 94--101, June 2018.

\bibitem{sutton1998reinforcement}
R.~S. Sutton, A.~G. Barto \emph{et~al.}, \emph{Reinforcement learning: An
  introduction}.\hskip 1em plus 0.5em minus 0.4em\relax MIT press, 1998-2018.

\bibitem{GozalvezCentralized}
D.~Calabuig, D.~Martin-Sacristan, M.~Botsov, J.~F. Monserrat, and D.~Gozalvez,
  ``{Comparison of LTE centralized RRM and IEEE 802.11 decentralized RRM for
  ITS cooperative awareness},'' in \emph{2018 IEEE Wireless Communications and
  Networking Conference (WCNC)}, April 2018, pp. 1--6.

\bibitem{cv2x11p}
V.~Vukadinovic, K.~Bakowski, P.~Marsch, I.~D. Garcia, H.~Xu, M.~Sybis,
  P.~Sroka, K.~Wesolowski, D.~Lister, and I.~Thibault, ``{3GPP C-V2X and IEEE
  802.11 p for Vehicle-to-Vehicle communications in highway platooning
  scenarios},'' \emph{Ad Hoc Networks}, vol.~74, pp. 17--29, 2018.

\bibitem{RLv2v}
H.~Ye and G.~Y. Li, ``{Deep reinforcement learning based distributed resource
  allocation for V2V broadcasting},'' in \emph{2018 14th International Wireless
  Communications Mobile Computing Conference (IWCMC)}, June 2018, pp. 440--445.

\bibitem{RLv2v2}
------, ``{Deep reinforcement learning for resource allocation in V2V
  communications},'' in \emph{2018 IEEE International Conference on
  Communications (ICC)}, May 2018, pp. 1--6.

\bibitem{RL11p}
A.~Pressas, Z.~Sheng, F.~Ali, D.~Tian, and M.~Nekovee, ``{Contention-based
  learning MAC protocol for broadcast vehicle-to-vehicle communication},'' in
  \emph{2017 IEEE Vehicular Networking Conference (VNC)}, Nov 2017, pp.
  263--270.

\bibitem{RLv2i}
R.~F. Atallah, C.~M. Assi, and M.~J. Khabbaz, ``Scheduling the operation of a
  connected vehicular network using deep reinforcement learning,'' \emph{IEEE
  Trans. on Intelligent Transportation Systems}, pp. 1--14, 2018.

\bibitem{sdn}
Q.~Zheng, K.~Zheng, H.~Zhang, and V.~C.~M. Leung, ``Delay-optimal virtualized
  radio resource scheduling in software-defined vehicular networks via
  stochastic learning,'' \emph{IEEE Transactions on Vehicular Technology},
  vol.~65, no.~10, pp. 7857--7867, Oct 2016.

\bibitem{cloud}
M.~A. Salahuddin, A.~Al-Fuqaha, and M.~Guizani, ``Reinforcement learning for
  resource provisioning in the vehicular cloud,'' \emph{IEEE Wireless
  Communications}, vol.~23, no.~4, pp. 128--135, August 2016.

\bibitem{Mnih2016}
V.~Mnih, A.~P. Badia, M.~Mirza, A.~Graves, T.~Harley, T.~P. Lillicrap,
  D.~Silver, and K.~Kavukcuoglu, ``Asynchronous methods for deep reinforcement
  learning,'' in \emph{Proceedings of the 33rd International Conference on
  International Conference on Machine Learning - Volume 48}, ser. ICML'16,
  2016.

\bibitem{nature}
Y.~LeCun, Y.~Bengio, and G.~Hinton, ``Deep learning,'' \emph{Nature}, vol. 521,
  no. 7553, p. 436, 2015.

\bibitem{ns3}
\BIBentryALTinterwordspacing
``{The ns-3 network simulator}.'' [Online]. Available: \url{www.nsnam.org}
\BIBentrySTDinterwordspacing

\bibitem{sumoPaper}
D.~Krajzewicz, J.~Erdmann, M.~Behrisch, and L.~Bieker, ``{Recent development
  and applications of SUMO-Simulation of Urban MObility},'' \emph{International
  Journal On Advances in Systems and Measurements}, vol.~5, no. 3\&4, 2012.

\bibitem{D2D}
R.~Rouil, F.~J. Cintr{\'o}n, A.~Ben~Mosbah, and S.~Gamboa, ``{Implementation
  and validation of an LTE D2D Model for ns-3},'' in \emph{Proceedings of the
  Workshop on ns-3}.\hskip 1em plus 0.5em minus 0.4em\relax ACM, 2017, pp.
  55--62.

\end{thebibliography}
